\documentclass{acm_proc_article-sp}

\usepackage{balance}  
\usepackage{graphics} 
\usepackage{url}      
\usepackage{graphicx,url}
\usepackage{array}
\usepackage[font=footnotesize]{subfig}
\usepackage{multirow}
\usepackage{bm}
\usepackage{hhline}
\usepackage{tabu}
\usepackage[T1]{fontenc}
\usepackage{times}
\usepackage{helvet}
\usepackage{courier}
\usepackage{datenumber}
\usepackage{hhline}

\newcounter{dateone}
\newcounter{datetwo}
\newcommand{\difftoday}[3]{%
      \setmydatenumber{dateone}{\the\year}{\the\month}{\the\day}%
      \setmydatenumber{datetwo}{#1}{#2}{#3}%
      \addtocounter{datetwo}{-\thedateone}%
      \the\numexpr(\thedatetwo - (-\thedatetwo/365)*365)\relax\space day(s)
} 
\newcommand{\superscript}[1]{\ensuremath{^{\textrm{#1}}}}

\def\wu{\superscript{*}}
\def\wg{\superscript{\dag}}
\begin{document}

\title{Mining the Minds of Customers from Online Chat Logs}

\CopyrightYear{2015}

\numberofauthors{1}
\author{
  \alignauthor Kunwoo Park\wu~~~~~~~Jaewoo Kim\wg~~~~~~~Jaram Park\wg~~~~~~~Meeyoung Cha\wg\\
    \affaddr{Graduate School of Web Science Technology, School of Computing\wu}\\
	\affaddr{Graduate School of Culture Technology\wg}\\    
    \affaddr{KAIST, South Korea}\\
    \email{\{kw.park,jaewoo.kim,jaram.park,meeyoungcha\}@kaist.ac.kr}\\
   \and
  \alignauthor Jiin Nam~~~~~~~~Seunghyun Yoon~~~~~~~~Eunhee Rhim\\
  	\affaddr{Intelligence Platform Lab, Software R\&D Center}\\
    \affaddr{Samsung Electronics, South Korea}\\
    \email{\{jiin.nam,sh001.yoon,eunhee.rhim\}@samsung.com}\\
}

\maketitle
\begin{abstract}
This study investigates factors that may determine satisfaction in customer service operations. We utilized more than 170,000 online chat sessions between customers and agents to identify characteristics of chat sessions that incurred dissatisfying experience. Quantitative data analysis suggests that sentiments or moods conveyed in online conversation are the most predictive factor of perceived satisfaction. Conversely, other session related meta data (such as that length, time of day, and response time) has a weaker correlation with user satisfaction. Knowing in advance what can predict satisfaction allows customer service staffs to identify potential weaknesses and improve the quality of service for better customer experience. 
\end{abstract}

\section{Introduction}

For businesses, operating Customer Service (CS) to recognize product-related problems and heighten customer satisfaction is a crucial management function~\cite{yan2004predicting}. Among the various strategies employed in traditional CS operations, much effort has been paid to efficiency and functionality such as providing prompt and accurate response to customers (e.g., minimize waiting times, optimize agent assignment)~\cite{kazmer2007identity,tan2013mediated}. 
Aspects of customer experience such as their moods and sentiments, on the other hand, have received relatively little attention. As many companies now employ live chats as a means of running their support service, this new method gives clear benefits over traditional phone support in terms of making available the logged chats immediately in text format for further analysis. 

Analyzing text logs in live chats can help identify what customers are saying about products and services as well as how well the support staff is performing, which is crucial for improving customer experience. Nonetheless, inferring customer feedback and satisfaction from text is not trivial. Existing methods rely on surveys posted to customers at the each support session, which call for voluntary participation. Another difficulty in customer experience is due to an asymmetric bimodal (J-shaped) distribution of feedback, where the average is a poor proxy of the overall quality. A large majority of customers often marking the service `very satisfactory' and another large group mark `very dissatisfying', which leads to divided opinions. 

Given the wealth of data in live chat systems, this research proposes a novel data mining approach to evaluate customer experience from unstructured chat dialogs. We explored the opportunity to mind the minds of customers based on a set of features about sessions and sentiments. By analyzing unstructured chat content, we can extract the possible feedbacks buried in chat logs and determine (1) what topics are being discussed in chat logs, (2) what sentiments those chats accompany, and (3) what the overall satisfaction level of customers is.

We utilized data of dyadic chat logs provided by the IT-mediated CS centers of Samsung Electronics, which offer a 24/7 live chat service (\textsf{http://www.samsung.com/us/support/live-chat.html})~\footnote{We plan to make an anonymized version of the chat data available for the wider research community. Please contact Eunhee Rhim for data if interested.}. \\The data are text-based and cover all conversation logs between customers and agents on the topic of mobile products of Samsung Electronics in the United States. The live chat system asks for feedback at the end of each chat session, yet only a marginal portion of customers (16\%) participate and a larger majority's experience (84\%) remain unknown. In order to infer customer satisfaction for these unevaluated sessions, we examine sentiments in text. This work was motivated by a finding in~\cite{shemwell1998customer} that says affective aspects of satisfaction are a key to a successful relationship between service providers and consumers.

\begin{figure*}[t!]
\minipage{0.32\textwidth}
  \includegraphics[width=\linewidth]{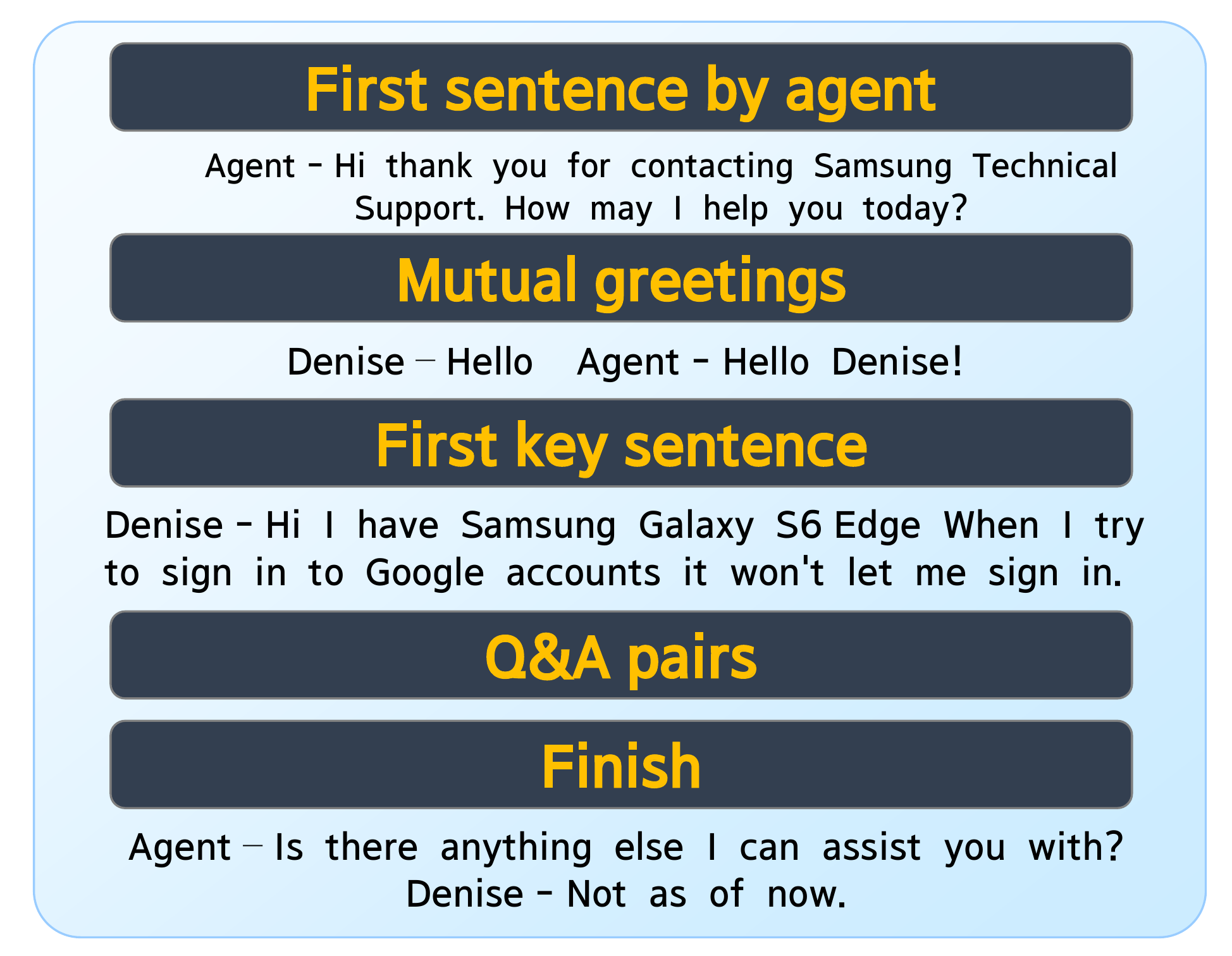}
  \hspace{-5mm}\centering
  \vspace{-2mm}
  \caption{Procedure of the Livechat service}\label{fig:flivechat}
\endminipage\hfill
\minipage{0.32\textwidth}
  \includegraphics[width=\linewidth]{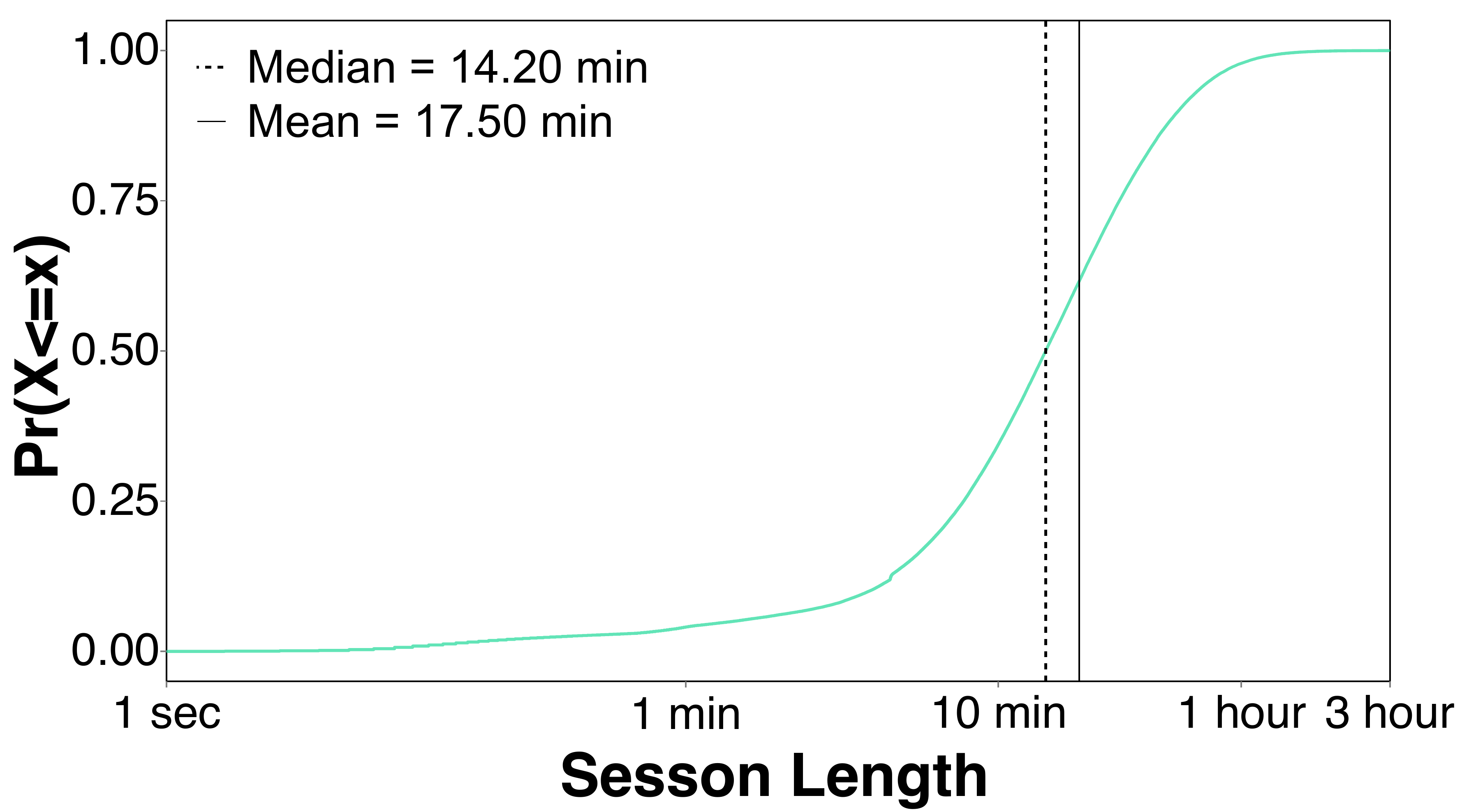}
  \caption{CDF on session lengths}\label{fig:session}
\endminipage\hfill
\minipage{0.32\textwidth}%
  \includegraphics[width=\linewidth]{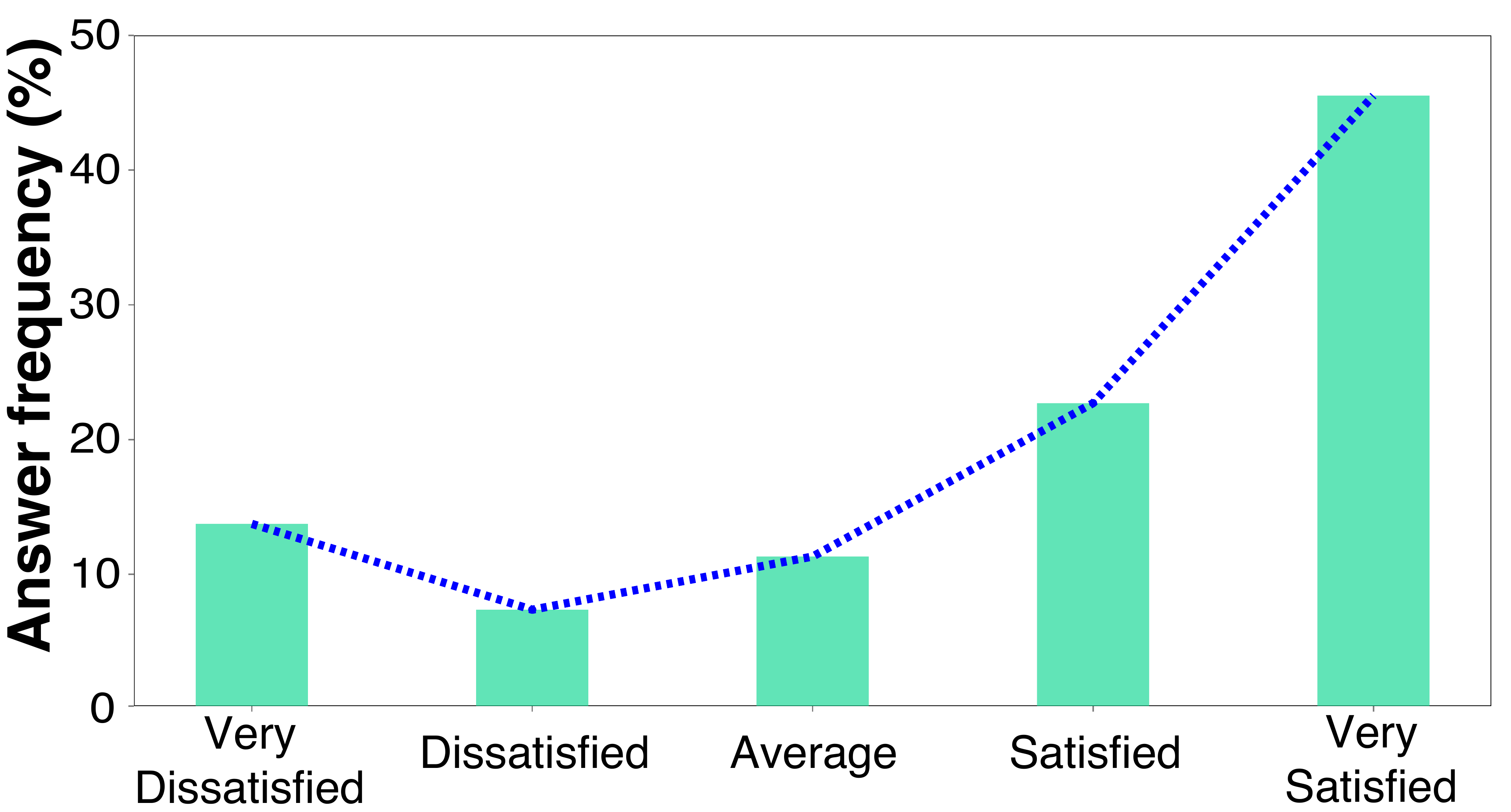}
  \caption{Post-chat survey results}\label{fig:satisfaction}
\endminipage
\end{figure*}

This work demonstrates how computational methods offer a chance to understand sentiments on conversational data through big data analysis. The key method relied on machine learning techniques and was cross-validated on pre-answered dataset. We make several findings. First, sentiments conveyed in online conversation is the most predictive factor of customer satisfaction. Conversely, other session related meta data (e.g., length, time of day) have a weaker correlation with user satisfaction. Second, this work identifies that not only sentiments of customers but also those from agents play an important role in predicting customer satisfaction. This finding suggests the feasibility of utilizing sentiments in inferring customer satisfaction. Inferring customer satisfaction from conversation could be an essential component for automated CS agent, which is expected to be realized in near future. Sentiments of agents were more stable over time than those of customers. We discuss a possible approach to expand sentiment dictionaries to better handle domain- and context-specific words for the task.

\section{Data Methodology}
\subsection{Live Chat Logs}

Upon entering Samsung Electronics's live chat system, customers are asked to choose the type of device they need help for (e.g., Galaxy S3). A chat session starts then as soon as an agent is allocated to that customer. Chats are in free text form, yielding varying intensity and length. Nonetheless, we could find a common structure by manual coding 3\% of sample sessions as depicted in Figure~\ref{fig:flivechat}. A chat is initiated by a greeting sentence from an agent (e.g., ``Hi, thank you for contacting Samsung Technical Support. How may I help you today?'') and is followed by mutual greetings. A customer typically describes a problem immediately after the greeting part, which is then followed by a series of question--and--answer--style conversation. In this work, we refer to each unit of conversation ``utterance''(i.e., the smallest unit in chats separated by pressing ``enter''), for which the log specifies the timestamp, the speaker, and text content. Often a given utterance is not a full sentence (e.g., ``Every time I try to sign into Google accounts''), when a sentence is spread across multiple utterance instances (e.g., ``I ran an update'' ``last week'' ``and I am not getting picture messages'' ``now'').

We obtained 12-month worth of complete conversation logs between customers and agents on a representative mobile products of Samsung Electronics in the United States. The dataset was given in the XML format and contained 173,886 sessions as well as 5,641,172 utterance instances. For each session, the log contained information about the following: (1) product type, (2) agent ID, (3) customer info (ID, geolocation, timezone), (4) duration info (start time, endtime, disconnecting entity), (5) utterance info (speaker, timestamp, text content), and (6) post-session survey results (4 questions).
Figure~\ref{fig:session} shows the distribution of session length, where the x-axis represents session length in a log scale. On average, each session lasts 17.50 minutes (Std=13.47), while the median is 14.20 minutes. This plot suggests that many sessions end within 20 minutes, while there exist extremely long sessions that lasted over one hour. 75\% of all sessions terminated after exchanging no more than 45 messages. 

Once a live chat session ends, customers are provided with survey questionnaires to indicate their satisfaction level as follows: 
\begin{enumerate}
\vspace*{-4mm}
\item \textit{How would you rate your overall satisfaction with the chat?} 

\vspace*{-2mm}
\item \textit{In the future, would you prefer to chat instead of calling?}

\vspace*{-2mm}
\item \textit{How would you rate satisfaction with the representative's overall knowledge? }

\vspace*{-2mm}
\item \textit{Please choose the main reason you were dissatisfied from the following choices.} 
\end{enumerate}
\vspace*{-4mm}

Answer for the first question represents the satisfaction score and was logged by 5-point Likert scale (Very Dissatisfied--Dissatisfied--Average--Satisfied--Very Satisfied). This survey helps us infer the quality of each session. Figure~\ref{fig:satisfaction} shows the distribution of responses, where the vertical axis represents the answer frequency. Note that these statistics are based on only 16\% of sessions that had complete survey results, as the survey was voluntary in nature. The overall satisfaction score follows a \textit{J}-shaped distribution, where a large fraction (45\%) found chats to be very satisfactory and another large group (14\%) indicated extreme unhappy experience. We later use this result to infer characteristics of sessions when customers were dissatisfied.

\subsection{Customer Sentiment Extraction}

\begin{figure*}[t!]
{
\centering
\includegraphics[width=0.85\linewidth]{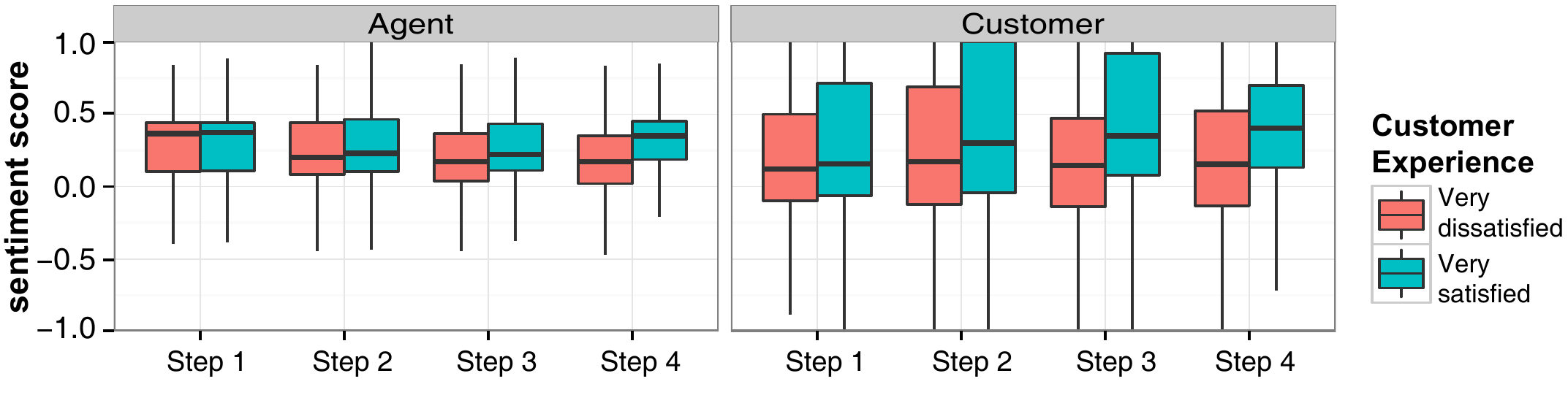}
 \caption{Dynamics of sentiment according to conversation stage}
 \label{fig:sentiment}
}
\end{figure*}

In order to identify sentiments from unstructured human chat data, we utilize an existing state-of-the-art sentiment tool called VADER (Valence Aware Dictionary and sEntiment Reasoner)~\cite{hutto2014vader}. VADER extends several human-validated sentiment lexicons such as LIWC, ANEW, and GL. One of the most popular sentiment analysis tool is LIWC (Linguistic Inquiry and Word Count), which does not perform well for online text containing slangs and short sentences. VADER constructed a gold-standard list of lexical features mainly focused on microblog context and performs well even for short messages. In this research, we analyzed sentiment of each utterance (i.e., the unit of chat dialog) through VADER. Examples chat utterances and their sentiment results are shown below. The score in parenthesis indicates valence (scaled from -1 to 1) a negative score refers to negative sentiment and vice versa.

\vspace*{-4mm}
\begin{quote}
\textit{Customer: I purchased phone and then I noticed a terrible scuff on my screen. (-0.389)}
\end{quote}
\vspace*{-5mm}
\begin{quote}
\textit{Agent: It was a pleasure assisting you, thank you for contacting Samsung Technical Support. (0.348)}
\end{quote}
\vspace*{-3mm}

As depicted in Figure~\ref{fig:flivechat}, chat dialogs usually followed common flow. Based on the flow, we may assume four steps of composition and we divide a conversation into an arbitrary set of four steps based on the total conversation length of each session as a simple approximation. We conducted sentiment analysis and aggregated the sentiment results for each conversation step. 

The resulting sentiment dynamics is shown in Figure~\ref{fig:sentiment} for two extreme sets of sessions: Very Satisfied (VS) group and Very Dissatisfied (VD) group. While the VS group exhibits more positive sentiment, the VD group shows more negative sentiment as conversation progresses for both customers and agents. It is interesting to observe that customers typically exhibit a wider variation in sentiments than agents. Both agents and customers start with a similar sentiment score. However, from Step 2 and onward, sentiments start to deviate between the two groups especially for customers. These changes imply customers' moods are dynamic, especially after the Step 1. Agent on the other hand have consistent sentiment in early stages (i.e., Step 1 and Step 2) regardless of customers' sentiment.

\section{Predicting Satisfaction}

We now seek to predict which sessions mark low satisfaction based on the available quantitative measures. We transformed the 5-point Likert Scale from user survey into a binary label to indicate whether each session is dissatisfactory to a customer or not. We did because identifying a dissatisfying session (rather than the opposite) is crucial for CS operation. Dissatisfied customers may be a minority group in terms of numbers, but meeting their needs is critical in customer management. Therefore we labelled satisfaction of customer as follows: (Very Dissatisfied, Dissatisfied) to TRUE, (Very Satisfied, Satisfied, Average) to FALSE in terms of \textit{dissatisfaction}. This grouping led to 4,649 TRUE and 20,175 FALSE sessions for prediction. To predict dissatisfaction, we extracted a total of 14 features for each chat steps across the following: session-level meta information (e.g., number of utterances, session length of a session), agent's sentiment (e.g., agent's sentiment at each step), and customer's sentiment (e.g., customer's sentiment at each step).

Table~\ref{table:performance} shows the performance of algorithms used for predicting dissatisfying sessions. Every measurement is based on 10-fold cross validation. For prediction, we utilize logistic regression, SVM with radial kernel (note as SVM-R), and random forest to classify sessions into binary scheme: TRUE/FALSE for dissatisfaction. As a baseline, we set up the majority voting and found accuracy of 0.8127. We make the following observations. Firstly, random forest outperforms the baseline. SVM-R is the best in terms of prediction accuracy, yet random forest yields the highest F1 score. Secondly, sentiments extracted from sessions can significantly improve the performance of the prediction task. Compared to the random forest model only using session-level meta information that is even beaten by the baseline (termed ``w/o sentiments''), incorporating sentiments makes a significant improvement in performance. Because label distribution is biased toward FALSE cases for dissatisfaction, it is naturally difficult to predict dissatsfaction well. However, the above results imply the importance of sentiments for understanding dissatisfying sessions.

{
\begin{table}[t] \small
  \centering \frenchspacing 
\begin{tabular}{c|ccc|c}
\hline
\multirow{3}{*}{Measure} & \multicolumn{3}{c|}{All features} & w/o sentiments\\\hhline{~----}
 &Logistic & \multirow{2}{*}{SVM-R} & Random & Random\\
 &Regression& & Forest & Forest\\\hline
Accuracy & 0.7941 & \textbf{0.8378} & 0.8372 & 0.8118 \\ 
F1 score & 0.3154 &0.2867 & \textbf{0.3949} & 0.0275 \\ 
\hline 
\end{tabular} 
\caption{Performances of algorithms for predicting customer's satisfaction under 10-fold cross validation}
\label{table:performance}
\vspace{-0.3cm}
\end{table}

}
\vspace{-0.3cm}

\begin{figure}[t]
{
\centering
\includegraphics[width=\linewidth]{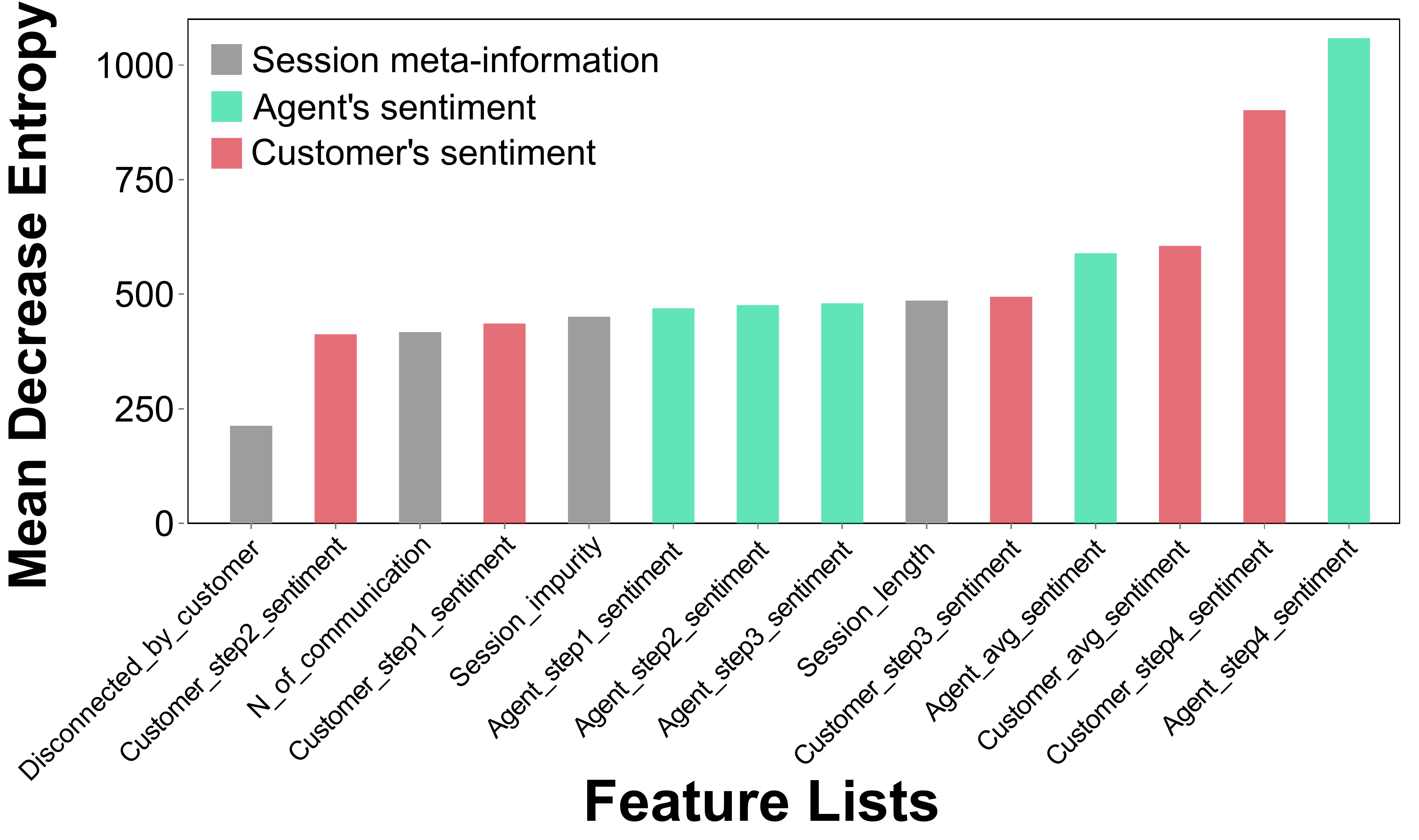}
 \caption{Feature importance for predicting customer's satisfaction}
 \label{fig:var_imp}
 \vspace{-0.3cm}
}
\end{figure}

Subsequently, we investigated the importance of individual features in the prediction task. Figure~\ref{fig:var_imp} shows the list of features and their mean decrease entropy in random forest. Entropy explains node impurity for a feature. Thus if a decision tree is well separated by a predictor, the measure will have a low value. Since there are multiple number of trees in a random forest, mean decrease entropy indicates the average importance of feature for each tree inside the random forest. The results show that sentiments play an important role in predicting customer's satisfaction for overall sessions. Sentiments of both customers and agents are important in understanding customer satisfaction. Moreover, agent's sentiment from Step 4 of the chat dialog is the most important feature in random forest, followed by the same quantity for customers. Agents, who generally are conservative in revealing their sentiments than customers, could express their sentiments stronger toward the end of chat sessions.

To figure out the context deeper, we conducted N-gram analysis for unigrams and bi-grams (N=1,2), and retrieved discriminative N-grams for dissatisfaction. Table~\ref{tab.chi.squ} presents the list of N-grams used in Step 4 of agent's conversation. In order to decide the discriminative power of each N-gram, we calculated Cramer's V, which is a normalized variant of Chi-squared value. Every term listed in this table is dominantly used in FALSE sessions (i.e., satisfactory or average). For example, `button' and `the blue' are more written in FALSE sessions than TRUE sessions (i.e., dissatisfying). Agents may ask customers to participate in post-survey by clicking a button (e.g., `please click', `button'). However in sessions where the customer's problem is not resolved properly, it may be difficult for agents to ask for a survey to customers or even customers may disconnect in the middle, leaving to a sudden end for that session.

\section{Discussion}

While this work demonstrated the possibility of applying sentiment analysis and machine learning to infer customer satisfaction, more efforts are needed before this idea can be implemented in a real system. Analyzing sentiments in dyadic chats between customers and agents is non-trivial for a number of reasons, because (1) individual variance is large, (2) sentiments tend to change over time, (3) utterances are unstructured and short in length, and (4) human sentiments are further subject to culture of the society they belong to. Chat data did not lend itself well to existing sentiment analysis tools in terms of coverage and biases---only 1.25\% of all utterances in chats contained any single word related to sentiments from state-of-the-art tools. Qualitative analysis that we conducted over a small subset of data revealed that automated tools sometimes could not understand sentiments and misjudged the valence. The following sentences contain negative sentiment based on a human coder's judgement, for instance, could not be understood correctly with automated sentiment analysis tools: 
\vspace{-0.4cm}
\begin{quote}
``\textit{I have tried to restart my phone over hundred times but the problem still persists}''\\\\
``\textit{My 5-year old can answer the same}''\\\\
``\textit{YEAH RIGHT}.''
\end{quote}
\vspace{-0.4cm}
In order to identify sentiments from unstructured human chat data, we plan to expand existing sentiment dictionaries to handle domain-specific words and expressions. For example, training with the CS chat data would allow us to fine tune the sentiment dictionary with domain- and context- specific words (e.g., `crack' would indicate problem on screens). Such joint expansion can be made through a variety of approaches including the PMI (Pointwise Mutual Information) approach~\cite{bross2013automatic}. The PMI-based framework assumes that when a word \textit{A} is frequently used with another word \textit{B} containing positive sentiment, \textit{A} is likely to represent positive sentiment as well. While omitted due to space limitation, we expended existing state-of-the-art sentiment dictionaries extensively and found potential for better understanding more complex sentences that often appear for the CS chat logs. For instance, a phrase such as `hundred times' in the above mentioned example becomes associated with negative sentiment in CS chats and hence the particular chat sentence can be better understood automatically. On top of the above approach, we could  adopt non-seed approaches~\cite{liang2014conr} to improve the coverage of sentiments in online chats.

\section{Conclusion}

The IT-based Customer Service (CS) is a crucial operation for a number of businesses around the world. In this study we tested the feasibility of utilizing chat text to infer customer satisfaction in CS conversation by utilizing computational sentiment extraction of conversational data. The main finding suggests that sentiments of both customers and agents play an important role for predicting customers' satisfaction, and consequently this can be used as a source of feedback to improve customer experience. Interestingly, agents tend to exhibit a lower degree of variance in their sentiments than customers, yet show negative sentiment at the last step of a chat particularly in those conversations that were marked as `very dissatisfying' (see Figure~\ref{fig:sentiment}). This result could imply that emotional interaction in the CS sessions is one of the key factors for understanding customer experience, so computational sentiment analysis could be a helpful albeit challenging tool. 

The sentiment lexicon we utilized is designed for understanding general online messages. For better performance, future studies can use CS domain-specified sentiment lexicons by expanding of existing sentiment dictionaries. In addition, service agents tend to ask customers to answer a post-chat survey when the conversation seems satisfactory to customers. This can cause a bias in answers and lead to a degree of skew to the \textit{J}-shaped distribution in survey result. Therefore, further investigation on distribution of unsatisfactory and satisfactory conversations could be conducted to identify a clear difference between them and improve prediction performance. 
{
\begin{table}[t] \small
\begin{tabular}{llll|lll}
\hline
\multirow{2}{*}{Rank} & \multirow{2}{*}{Unigram} & \multicolumn{2}{c|}{Frequency}&\multirow{2}{*}{Bigram} & \multicolumn{2}{c}{Frequency} \\

& & TRUE  & FALSE & & TRUE & FALSE \\
\hline
1    & button      & 0.294  & 0.801 & the blue      & 0.280 & 0.774 \\
2    & survey      & 0.294  & 0.796 & transcript of & 0.275 & 0.758 \\
3    & blue        & 0.282  & 0.776 & please click  & 0.273 & 0.755 \\
4    & fill        & 0.280  & 0.767 & to receive    & 0.282 & 0.766 \\
5    & transcript  & 0.288  & 0.791 & your chat     & 0.273 & 0.751 \\
6    & pleasure    & 0.275  & 0.775 & button to     & 0.272 & 0.749 \\
7    & click       & 0.300  & 0.791 & a transcript  & 0.273 & 0.749 \\
8    & close       & 0.271  & 0.747 & Have a        & 0.288 & 0.811 \\
9    & assisting   & 0.269  & 0.744 & It was        & 0.275 & 0.775 \\
10   & brief       & 0.271  & 0.744 & receive a     & 0.274 & 0.752 \\
\hline
\end{tabular}
\caption{Discriminative N-grams (N=1,2) ranked by Cramer's V in the 4th step of session for agents}
\label{tab.chi.squ}
 \vspace{-0.3cm}
\end{table}
}

{

\bibliographystyle{abbrv}
\bibliography{references}  
}

\balancecolumns
\end{document}